# A General protocol for Distributed Quantum Gates

Moein Sarvaghad-Moghaddam[1], Mariam Zomorodi-Moghadam[2,*], Ahmed Farouk[3]

[1] Quantum Design Automation Lab, Amirkabir University of Technology, Tehran, Iran.

[2] Department of Computer Engineering, Ferdowsi University of Mashhad, Mashhad, Iran.

[3] Department of Physics and Computer Science, Wilfrid Laurier University, Waterloo, Canada.

**Abstract** Distributed quantum computation requires to apply quantum remote gates on separate nodes or subsystems of network. On the other hand, Toffoli gate is a universal and well-known quantum gate. It is frequently used in synthesis of quantum circuits. In this paper, a general protocol for implementing a remote $n$-qubit controlled-$U$ gate is presented with minimum required resources. Then, the proposed method is applied for implementing a Toffoli gate in bipartite and tripartite systems. This method also is optimal when group of the qubits belong to one part, section or subsystem of network. Beucase it only use one entangled qubit for each group of the qubits in mentioned conditions.

## 1    Introduction

Interest in quantum computing has increased with great potential in solving specific problems and it is becoming an important computational issue [1-7]. The theory of quantum computing is getting more and more mature since it was initiated by Feynman and Deutsch in the 1980s [8,9]. Compared with classical computing, quantum computing has the outstanding advantages in terms of the speed of computing. Quantum computation has revolutionized computer science, showing that the processing of quantum states can lead to a tremendous speed up in the solution of a class of problems, as compared to traditional algorithms that process classical bits [2,3].

A large-scale quantum computer is needed to solve complex problems at higher speeds. But, there are some problems in implementation of a large-scale quantum system. Due to the interaction of qubits with the environment that leads to quantum decoherence and more sensitivity to errors [10-12],the number of qubits used in processing information should be limited. One reasonable solution for overcoming to the mentioned problem is distributed quantum computer. A distributed quantum computation can be built using two or more low-capacity quantum computers with fewer qubits as distributed nodes or subsystems in a network of quantum system for solving a single problem [13,14]. Distributed quantum computation first had been proposed by Grover [15], Cleve and Buhrman [16], and Cirac et al. [17]. Then, Ying and Feng [11] defined an algebraic language for describing a distributed quantum circuits. After that, Van Meter et al. [18] proposed a structure for VBE carry-ripple adder in a distributed quantum circuit.

One the other hand, to setup a distributed quantum system, a communication protocol is needed between its separate nodes. In 2001, Yepez [19] proposed idea using of classical communication instead of quantum communication in interconnecting the subsystems or nodes of distributed quantum computers called as Type-II quantum computers. In this paper, quantum communication (type-I) is used for interconnecting the subsystems of a distributed quantum computer. One of methods for transmitting qubits with unconditional security, between nodes of network is Quantum Teleportation (QT) [20-23]. In teleportation, qubits are transmitted between two users or nodes, without physically moving them and then computations are locally performed on qubits, which is also known as teledata. There is an alternative approach, called as telegate that executes gates remotely and directly using the quantum entanglement when nodes are in a long distance. One of the problems in the second method is to establish optimal implementations of quantum gates between qubits that are located in different nodes of the distributed quantum computer. One of well-known reversible and quantum gates is Toffoli gate that is universal. I.e. any reversible and quantum circuit can be constructed from Toffoli gates. So, it is important to implementation of a protocol for applying n-qubit remote Toffoli gate between separate nodes of network.

---

* Corresponding Author.
Email: m_zomorodi@um.ac.ir

Moein Sarvaghad-Moghaddam
Email: moeinsarvaghad@yahoo.com

In recent decades, only, a few papers [24-26] has focused on implementing remote quantum gates. paper [24] presented a method for implementing remote n-qubit Toffoli gate. Also in [25] a method for implementation of remote three-qubit Toffoli gate using of an auxiliary 4-dimensional quantum system is presented. Then, using synthesizing a $n$-qubit Toffoli gate to three-qubit Toffoli gates, an implementation for $n$-qubit remote Toffoli gate is presented. These methods are not optimal when there are some of qubits in one node or subsystem of network.

In this paper, we presented a general protocol for implementing of remote n-controlled-U gate using minimum resources. In comparing with previous works, this method is especially optimal when many of qubits belong to one part, section or subsystem. in other cases, the proposed method is same with previous works in used resources.

The rest of the paper is organized as follows: Section 2 present some related background about quantum computation and distributed quantum circuits. In Section 3 the proposed method is introduced in details. Section 4 present a comparison and discussion about the proposed method. Finally, Section 5 concludes the paper.

## 2  Basic Concepts

Quantum states can be represented by vectors or a more famous notation of bra/Ket. Kets (shown as $|x\rangle$) display column vectors and are generally used to describe quantum states. The bra notation (shown as $\langle x|$) displays transpose conjugate of $x$ vector ($|x\rangle$). Basic states of $|1\rangle$ and $|0\rangle$ can be stated as vectors of $[0\ 1]^T$ and $[1\ 0]^T$ respectively. Any combination of $|1\rangle$ and $|0\rangle$ states ($\alpha|0\rangle + \beta|1\rangle$) can be showed as $[\alpha\ \beta]^T \epsilon\ C^2$, in which $C$ denote set of complex number.

A qubit is a unit vector in a complex two-dimensional space that the specific basis vectors with the notation of $|0\rangle$ and $|1\rangle$ have been selected for this space. The base vectors of $|0\rangle$ and $|1\rangle$ are quantum counterparts of classic bits of 0 and 1, respectively. Unlike classical bits, qubits can be in any superposition of $|0\rangle$ and $|1\rangle$ like $\alpha|0\rangle + \beta|1\rangle$ where α and β are the complex numbers such that $|\alpha|^2 + |b|^2 = 1$. If such a superposition is measured compare with the base of $|0\rangle$ and $|1\rangle$, then $|0\rangle$ and $|1\rangle$ are observed with probability of $|\alpha|^2$ and $|\beta|^2$, respectively.

### 2.1  Quantum and reversible gates

Quantum logic is inherently reversible [27]. Quantum operations can be achieved with a network of gates. Each quantum gate is a linear transformation that is defined on the $n$-qubit Hilbert space by an effective unitary matrix. The matrix $U$ is unitary, if $UU^\dagger = 1$ where $U^\dagger$ is the transpose conjugate of matrix $U$. In the following, a definition of well-known quantum and reversible gates used in the proposed method is presented. The first, a $n$-qubit Toffoli gate [28] can be defined as the form $TOF(C, T)$, where $C = \{x_{i_1}, x_{i_2}, \dots, x_{i_m}\} \subset X$ is set of control lines and $T = \{x_j\}$ with $C \cap T = 0$ is the target line. If all the controls have value 1, the target line is inverted; otherwise the value on the target line is passed through unchanged. For $m = 0, m = 1$, and $m = 2$ the gates are called $NOT$, $CNOT$, and $C^2NOT$ ($Toffoli$) respectively. Generally, above definition can be used for each $n$-qubit controlled gate. Another operation used in the proposed method is Hadamard, $Z$ ($\sigma_Z$) and $X$ operations that can be defined as the following:

$$H = \frac{1}{\sqrt{2}}\begin{bmatrix}1 & 1\\1 & -1\end{bmatrix}, Z = \begin{bmatrix}1 & 0\\0 & -1\end{bmatrix}, X = \begin{bmatrix}0 & 1\\1 & 0\end{bmatrix} \qquad (1)$$

Fig. 1. a)- d) show schematics of Hadamard, CNOT, Toffoli, and controlled-Z, respectively.

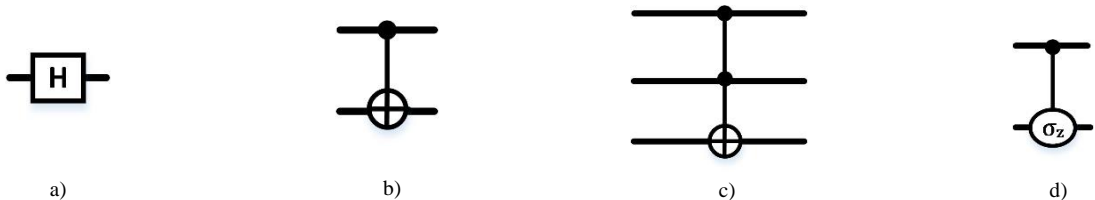

Fig. 1: Schematic of  a) Hadamard, b) CNOT, c) Toffoli, and d) controlled-Z

### 2.2  Distributed Quantum Computation

A *Distributed Quantum Circuit (DQC)* is a network consist of limited capacity *Quantum Circuits (QCs)* which are as partitions or nodes of network. This nodes are located in a long distance from each other and overall emulate the functionality of a large quantum system. This nodes are connected together via a specific quantum communication channel such as teleportation or a classical channel for communicating partitions (nodes) of $DQC$ by sending their qubits or their measurement results to each other. Qubits are numbered from top to bottom from one to $n$ in each partition, where the $i_{th}$ line of the circuit from top to bottom represents the $i_{th}$ qubit, $q_i$. In quantum information science we think of entanglement, and especially distributed entanglement, as a resource that is useful for tasks such as quantum key distribution, teleportation, and distributed quantum computation.

There are two kinds of quantum gates in the $DQC$, namely, local and global gates. In local gate, controls and target qubits belong to the same partition, node or subsystem while control and target qubits belong to different partitions, nodes or subsystems in a global gate.

In the next section our proposed method for implementing an n-qubit controlled-$U$ gate in a distributed quantum circuit is described. This gate can be used in a distributed quantum system.

## 3  Proposed Protocol

In this section, a novel protocol is presented for implementing a global (remote) $n$-qubit controlled-$U$ gate in a distributed quantum circuit. Then the protocol is applied especially for implementing a Toffoli gate. Fig. 2.a shows the proposed protocol for an $n$-qubit controlled-$U$ gate distributed between three remote nodes, parts, or subsystem, A, B and C. But, generally, it can be generalized and used with more parts, nodes or subsystems. As shown in Fig. 2.b, control lines of corresponding $U$ gate are qubits $A_1, \ldots, A_n, B_1, \ldots, B_n, C_1, \ldots, C_{n-1}$ and target line is $C_n$.

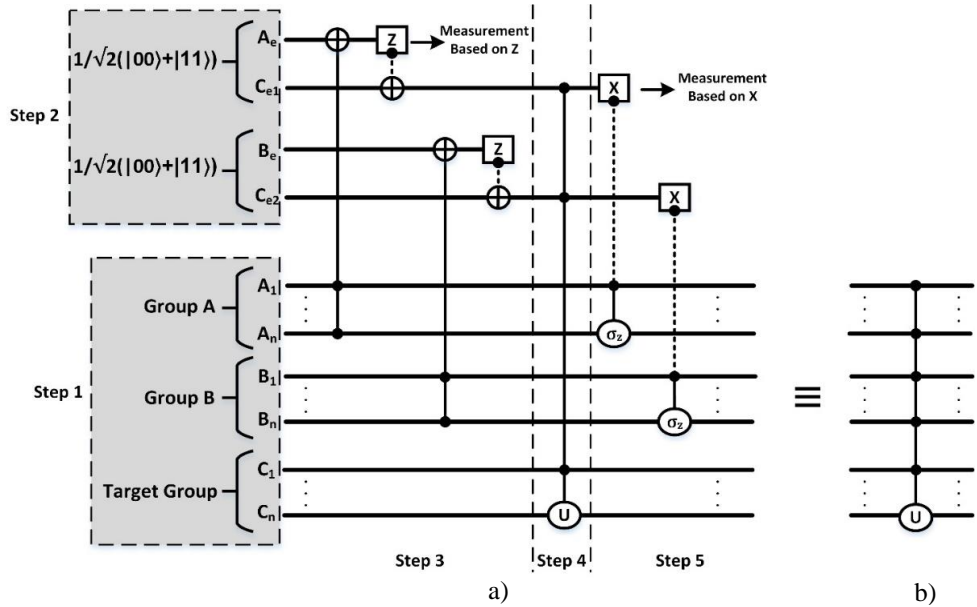

Fig. 2: a) Implementation of global $n$-qubit controlled- $U$ gate. b) Schematic of corresponding $n$-qubit controlled-$U$ gate.

The steps of the proposed protocol are as the following:

**Step 1.** If many qubits exist in possession of one node, section, or subsystem then, they are placed in one group. Especially, if target line exists in one section, it is called target group. For example, Fig. 2.b) shows a remote Toffoli gate in possession of three section A, B and C that it can be converted to three groups include of Group A, Group B and Target Group.

**Step 2.** For each group that is distinct from target group, an entangled state (Bell state) is added belong to this group and target group. For example, in Fig. 2, there is two distinct parts (Group A and Group B) and Target Group. So, it need to two dependent Bell state between Group A and Target Group (qubits $A_e$ and $C_{e1}$) and Also, Group B and Target Group ($B_e$ and $C_{e2}$).

**Step 3**. A Toffoli gate is applied between each group as control line and its corresponding entangled qubit as target line. Then, a basis- $Z$ measurement is applied on corresponding entangled qubit and then its result is transmitted to another entangled qubit that is belong to target group, on classical channel and if its result is $|1\rangle$ a NOT ($X$) gate is applied in corresponding target group. For example, in Fig. 2, a Toffoli gate is applied between qubits $A_1$ to $A_n$ as control line and $A_e$ as target line. Afterwards a measurement based on Z on qubit $A_e$ is applied and its result is transmitted to target group with classical channel. then a CNOT gate is applied so that the measurement result is as control qubit and qubit $C_{e1}$ as target qubit. This action is done for other groups (except target group).

**Step 4.** A controlled-$U$ gate is applied between all qubits belong to Target Group as control lines and main target qubit as target line. For example, in Fig. 2, a controlled-$U$ gate is applied to qubits $C_{e1}, C_{e2}, C_1, \ldots C_{n-1}$ as control line and $C_n$ as target line.

**Step 5.** A measurement based on $X$ is applied to entangled qubits belong to Target group and then results are transmitted to dependent groups to entangled state. Then, according to this results, a controlled-$Z$ is applied to qubits in dependent groups. For example, in Fig. 2, a measurement based on X is applied to qubit $C_{e1}$ ($C_{e2}$), then results are transmitted to dependent groups i.e. Group A (Group B). If results are $|-\rangle$, then a controlled-$Z$ gate is applied to qubits $A_1$ to $A_{n-1}$ ($B_1$ to $B_{n-1}$) as control line and $A_n$ ($B_n$) as target line. Finally corresponding remote Toffoli gate is applied and the protocol is successfully finished.

In the following, two special case of global Toffoli gate is explained using the proposed protocol in details.

### a. Toffoli Gate on a Bipartite System

Suppose qubit $A_1$ ($A_1, A_2$) belong to Alice and qubits $B_1, B_2$ ($B_1$) belong to Bob as shown in Fig. 3.a) (Fig. 3.b)). So, in this case, there are two section, node or subsystem, group A and target group. For applying remote Toffoli gate between Alice and Bob so that qubits $A_1$ and $B_1$ ($A_1$ and $A_2$) are control qubits and target is $B_2$ ($B_1$), two qubits $A$ and $B$ are defined as maximally entangled state as a quantum channel between Alice and Bob as the following:

$$|\phi\rangle_{AB} = \frac{1}{\sqrt{2}}(|00\rangle + |11\rangle) \tag{1}$$

Fig. 3.a) (Fig. 3.b)) show proposed protocol for implementing global Toffoli gate on qubits $A_1$ and $B_1$ ($A_1$ and $A_2$) as control inputs and $B_2$ ($B_1$) as target for first case (second case).

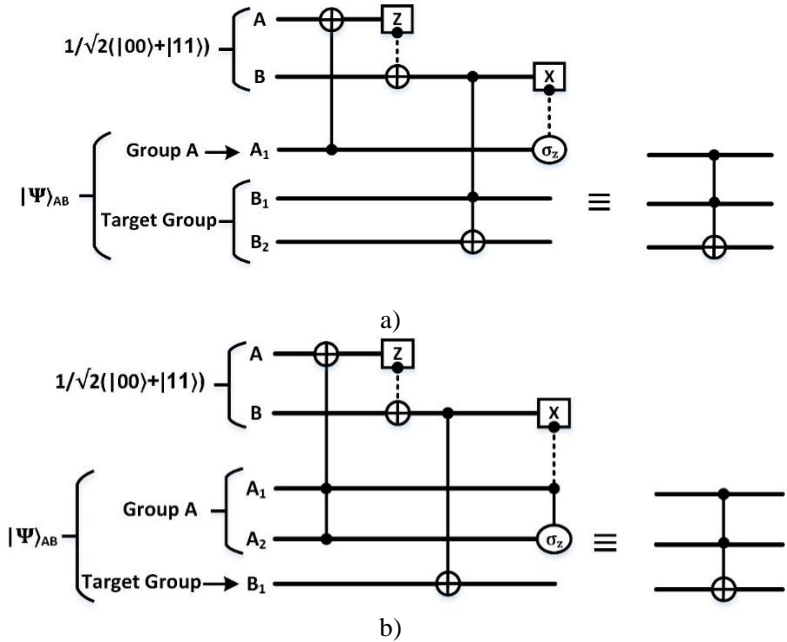

Fig. 3: Illustration of five-qubit circuit used for quantum teleportation of Toffoli gate in bipartite system a) first case b) second case.

Then, the input state of the qubits $A_1, B_1$ and $B_2$ ($A_1, A_2$ and $B_1$) can be expressed in arbitrary general state as Eq. (2).

$$|\Psi\rangle_{AB} = d_0|000\rangle + d_1|001\rangle + d_2|010\rangle + d_3|011\rangle + d_4|100\rangle \\ + d_5|101\rangle + d_6|110\rangle + d_7|111\rangle \tag{2}$$

The general state of system is as Eq. (3)

$$|\psi\rangle = |\Psi\rangle_{AB} \otimes |\phi\rangle_{AB} = \frac{1}{\sqrt{2}} (d_0|000\rangle + d_1|001\rangle + d_2|010\rangle + d_3|011\rangle + d_4|100\rangle + d_5|101\rangle + d_6|110\rangle + d_7|111\rangle) \otimes (|00\rangle + |11\rangle) \quad (3)$$

Then, the following steps are applied:

**Step 1.** Controlled-NOT gate (Toffoli gate) is applied so that $A_1$ ($A_1$ and $A_2$) is as control qubit and $A$ is target. The next state of system is as Eq. (4) (Eq. (5)).

$$CNOT_{A1,A}|\Psi\rangle = \frac{1}{\sqrt{2}} [(d_0|000\rangle + d_1|001\rangle + d_2|010\rangle + d_3|011\rangle)|00\rangle + (d_4|100\rangle + d_5|101\rangle + d_6|110\rangle + d_7|111\rangle)|10\rangle + (d_0|000\rangle + d_1|001\rangle + d_2|010\rangle + d_3|011\rangle)|11\rangle + (d_4|100\rangle + d_5|101\rangle + d_6|110\rangle + d_7|111\rangle)|01\rangle]_{A_1B_1B_2AB} \quad (4)$$

$$TOF_{A_1,A_2,A}|\Psi\rangle = \frac{1}{\sqrt{2}} [(d_0|000\rangle + d_1|001\rangle + d_2|010\rangle + d_3|011\rangle + d_4|100\rangle + d_5|101\rangle)(|00\rangle + |11\rangle) + (d_6|110\rangle + d_7|111\rangle)(|10\rangle + |01\rangle)]_{A_1A_2B_1AB} \quad (5)$$

**Step 2.** Alice apply single-qubit measurement on her qubit $A$ in $Z$-basis. Then, according to that, she notify to Bob her result and apply $X$ operation to his qubits so that if Alice's measurement result be one, Bob apply $X$ operation to qubit $B$.

**Step 3.** In this step, Bob apply a local Toffoli gate (CNOT gate) between qubits $B_1, B$ ($B$) as control qubits and $B_2$ ($B_1$) as target so that if qubits $B_1$ and $B$ ($B$) be one then, target $B_2$ ($B_1$) is one. Afterwards, the general state of system is as Eq. (6) (Eq. (7)).

$$TOF(B_1, B, B_2)CNOT_{A1,A}|\Psi\rangle = \frac{1}{\sqrt{2}} [(d_0|000\rangle + d_1|001\rangle + d_2|010\rangle + d_3|011\rangle)|0\rangle + (d_4|100\rangle + d_5|101\rangle + d_6|111\rangle + d_7|110\rangle)|1\rangle]_{A_1B_1B_2B} \quad (6)$$

$$\frac{1}{\sqrt{2}} [(d_0|000\rangle + d_1|001\rangle + d_2|010\rangle + d_3|011\rangle + d_4|100\rangle + d_5|101\rangle)(|0\rangle) + (d_6|111\rangle + d_7|110\rangle)(|1\rangle)]_{A_1A_2B_1B} \quad (7)$$

**Step 4.** After applying Toffoli gate (CNOT gate), Bob, apply single-qubit measurement on his qubit $B$ in $X$-basis. Then, he notify his measurement result to Alice. Afterwards she apply $Z$ (controlled-$Z$) operation according to Bob's measurement result. If his measurement result be $|-\rangle$, she apply $Z$ (controlled-$Z$) operation to $A_1$ ($A_1$ and $A_2$). Then, the general state of system is as Eq. (6) and the protocol is successfully finished.

$$d_0|000\rangle + d_1|001\rangle + d_2|010\rangle + d_3|011\rangle + d_4|100\rangle + d_5|101\rangle + d_6|111\rangle + d_7|110\rangle \quad (6)$$

### b. Toffoli Gate on a Tripartite System

In this section, as shown in Fig. 4, we suppose the global Toffoli is so that two control qubit ($A_1$ and $B_1$) are belong to Alice and Bob, respectively; and target is belong to Charlie ($C$). So, in this structure, there is three Group A, B and Target Group. For this aim, we need to a four qubit entangled state as a quantum channel so that qubits $A$ and $B$ belong to Alice and Bob and also, qubits $C_1$ and $C_2$ belong to Charlie. Quantum Channel created can be described as Eq. (7).

$$|\phi\rangle_{ABC} = \frac{1}{\sqrt{2}}(|00\rangle + |11\rangle)_{AC_1} \otimes \frac{1}{\sqrt{2}}(|00\rangle + |11\rangle)_{BC_2} \quad (7)$$

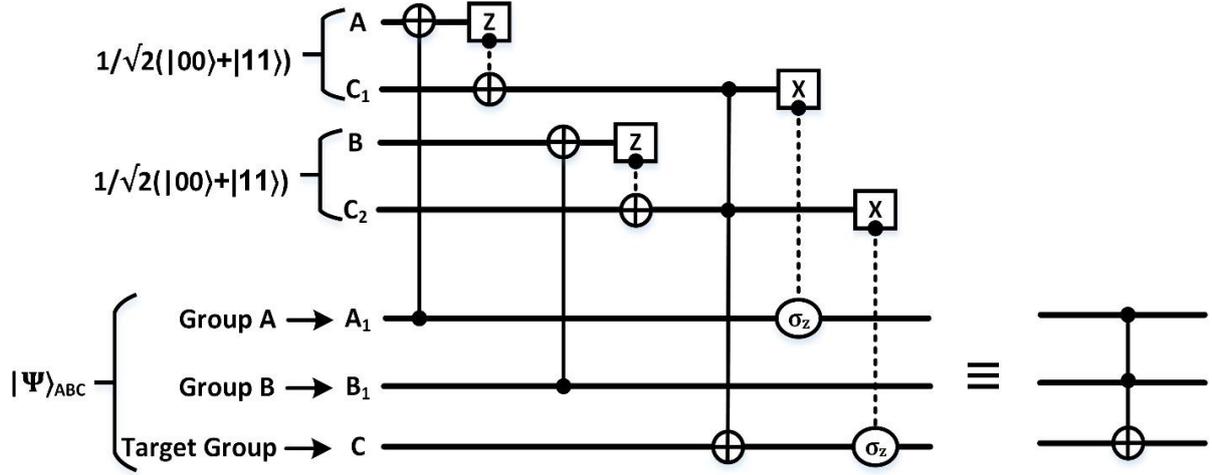

Fig. 4: Illustration of quantum teleportation of Toffoli gate in tripartite system.

Then, the initial state of the arbitrary qubits $A_1, B_1$ and $C$ can be stated as general Eq. (8).

$$|\Psi\rangle_{ABC} = (d_0|000\rangle + d_1|001\rangle + d_2|010\rangle + d_3|011\rangle + d_4|100\rangle + d_5|101\rangle + d_6|110\rangle + d_7|111\rangle)_{A_1 B_1 C} \quad (8)$$

The general state of system is as Eq. (9).

$$|\psi\rangle = |\Psi\rangle_{ABC} \otimes |\phi\rangle_{ABC} = \frac{1}{2}(d_0|000\rangle + d_1|001\rangle + d_2|010\rangle + d_3|011\rangle + d_4|100\rangle + d_5|101\rangle + d_6|110\rangle + d_7|111\rangle)_{A_1 B_1 C} \otimes (|00\rangle + |11\rangle)_{AC_1} \otimes (|00\rangle + |11\rangle)_{BC_2} \quad (9)$$

Then, the following steps are applied:

**Step 1.** CNOT gates are applied so that $A_1$ and $B_1$ are as control qubits $A$ and $B$ are targets, respectively. The general state of system is as (10).

$$CNOT_{A_1,A} CNOT_{B_1,B} |\psi\rangle = \frac{1}{2}[(d_0|000\rangle + d_1|001\rangle)(|00\rangle + |11\rangle)(|00\rangle + |11\rangle) + (d_2|010\rangle + d_3|011\rangle)(|00\rangle + |11\rangle)(|10\rangle + |01\rangle) + (d_4|100\rangle + d_5|101\rangle)(|10\rangle + |01\rangle)(|00\rangle + |11\rangle) + (d_6|110\rangle + d_7|111\rangle)(|10\rangle + |01\rangle)(|10\rangle + |01\rangle)]_{A_1 B_1 C A C_1 B C_2} \quad (10)$$

**Step 2.** In this step, Alice and Bob apply single-qubit measurements on their qubits $A$ and $B$ in $Z$-basis, and notify your results to Charlie. Then, Charlie apply controlled-NOT operations to his qubits $C_1$ and $C_2$ according to their measurement results so that if measurement result is $|1\rangle$, then he apply $X$ operation to corresponding qubits. The general state of system is as (11).

$$\frac{1}{2}[(d_0|000\rangle + d_1|001\rangle)|00\rangle + (d_2|010\rangle + d_3|011\rangle)|01\rangle + (d_4|100\rangle + d_5|101\rangle)|10\rangle + (d_6|110\rangle + d_7|111\rangle)|11\rangle]_{A_1 B_1 C C_1 C_2} \quad (11)$$

**Step 3.** Charlie apply a local toffoli gate between qubits $C_1$ and $C_2$ as control qubits and $C$ as target. The next state is as (12).

$$\frac{1}{2}[(d_0|000\rangle + d_1|001\rangle)|00\rangle + (d_2|010\rangle + d_3|011\rangle)|01\rangle + (d_4|100\rangle + d_5|101\rangle)|10\rangle + (d_6|111\rangle + d_7|110\rangle)|11\rangle]_{A_1 B_1 C C_1 C_2}$$

(12)

**Step 4**. Charlie apply a $X$-basis measurement on his qubits $C_1$ and $C_2$. Then he notify results to Alice and Bob, respectively. According to her results, Alice and Bob apply a $Z$ operation when her result be $|-\rangle$. The final state of system is the following and the protocol is successfully finished.

$$\frac{1}{\sqrt{2}} [(d_0|000\rangle + d_1|001\rangle + d_2|010\rangle + d_3|011\rangle + d_4|100\rangle + d_5|101\rangle + d_6|111\rangle + d_7|110\rangle]_{A_1 B_1 C} \quad (13)$$

## 4 Comparison and Discussion

In this section, a comparison is presented between the proposed protocol and the previous works [25,24] as shown in Table 1. In this table, for simplicity, especially consider bipartite and tripartite Toffoli gate in previous section in terms of the number of entangled gates, the number of auxiliary qubits, the number of applied operations and the number of measurements. As shown in this table, CE, SM and FM stand for controlled elevation as introduced in [25], single-qubit measurement and four-qubit measurement, respectively. In [25], a method is presented for applying remote Toffoli gate using an auxiliary four-dimensional quantum system. Then for generalizing this method, a $n$-qubit toffoli gate is converted to many Toffoli gate and this method is applied on it. In [24], a method is proposed for applying $n$-qubit Toffoli gate. In this method, an entangled-qubit is considered for transmitting each qubit. So, these methods use additional resources and are not optimal. In this paper, we present a method for applying $n$-qubit controlled-$U$ gate using division of qubit's ownership to groups. So, in this method, result of each group is transmitted to target group. In this regard, the proposed method is optimal into previous ones especially when we need to apply a $n$-qubit Toffoli gate with group of qubits in one subsystem, node or part. Because this method only use one entangled qubit. For example, consider scenario proposed in Fig. 1. This figure shows a remote Toffoli gate with $3n$ inputs so that $n$-inputs belong to Alice, $n$-inputs belong to Bob and $n$-inputs belong to Charlie. Table 2, show a comparison between the proposed method and [24] for this scenario. As shown in this table, our method has used fewer resources compared previous ones in terms of number of entangled resources, applied operations and measurements.

Table 1: Comparison of the proposed method with methods [24,25] for a bipartite and tripartite Toffoli gate.

| Type | Method | #NUM Entangled qubits | #Num auxiliary qubits | #NUM applied operations | #NUM measurements |
|---|---|---|---|---|---|
| Bipartite: Case 1 | [25] | 1 | 4 | 3 CE, 2 CNOT | 2 SM, 1 FM |
| | [24] | 1 | 0 | 1 CNOT, 1 Toffoli, 1 Hadamard | 2 SM |
| | The proposed protocol | 1 | 0 | 1 CNOT, 1 Toffoli | 2 SM |
| Bipartite: Case 2 | [25] | 1 | 4 | 3 CE, 2 CNOT | 2 SM, 1 FM |
| | [24] | 2 | 0 | 2 CNOT, 1 Toffoli, 2 Hadamard | 4 SM |
| | The proposed protocol | 1 | 0 | 1 Toffoli, 1 CNOT | 2 SM |
| Tripartite | [25] | 4 | 4 | 3 CE, 3CNOT | 3 SM, 1 FM |
| | [24] | 2 | 0 | 2 CNOT, 1 Toffoli, 2 Hadamard | 4 SM |
| | The proposed protocol | 2 | 0 | 2 CNOT, 1 Toffoli | 4 SM |

Table 2: Comparison of the proposed method with method [24] for scenario proposed in Fig. 2.

| Method | #NUM Entangled qubits | #Num auxiliary qubits | #NUM applied operations | #NUM measurements |
|---|---|---|---|---|
| [24] | $3n-1$ | 0 | $(3n-1)$ CNOT<br>1 $3n$-qubit Toffoli<br>$(3n-1)$ Hadamard | $2(3n-1)$ |
| The proposed protocol | 2 | 0 | 2 $(n+1)$-qubit Toffoli<br>1 $(n+2)$-qubit Toffoli | 4 |

## 5 Conclusion

In this paper, a general protocol for implementing a remote n-qubit controlled-U gate was presented with minimum required resources. Then, the proposed method was applied for implementing a Toffoli gate in bipartite and tripartite systems. This method was also optimal when many of qubits belong to one part, section or subsystem of network.